\documentclass[12pt]{article}

\usepackage{amsmath,amssymb,amsfonts,graphics,graphicx,amscd,amsfonts,epsfig,color}
\usepackage[hidelinks,pdftex]{hyperref}
\usepackage[font=small,labelfont=small]{caption}
\newcommand{\su}[1]{{\rm SU}(#1)}

\allowdisplaybreaks[4]

\date{}
\begin{document}

\begin{flushright}

\end{flushright}

\vspace{0.1cm}

\begin{center}

   {\Large
Anatomy of Deconfinement
  }   
   
\end{center}
\vspace{0.1cm}
\vspace{0.1cm}

\begin{center}
Masanori Hanada$^a$, Antal Jevicki$^b$, Cheng Peng$^{b,c}$ and Nico Wintergerst$^d$
\vspace{0.5cm}

$^a${\it School of Physics and Astronomy, and STAG Research Centre}\\
{\it University of Southampton, Southampton, SO17 1BJ, UK}\\
\vspace{0.2cm}  
$^b${\it Department of Physics, Brown University, 182 Hope Street, Providence, RI 02912, USA}\\
\vspace{0.2cm}  
$^c${\it Center for Quantum Mathematics and Physics (QMAP), Department of Physics\\
University of California, Davis, CA 95616 USA}\\
\vspace{0.2cm}  
$^d${\it The Niels Bohr Institute, University of Copenhagen,\\ 
Blegdamsvej 17, 2100 Copenhagen \O, Denmark}
\end{center}

\vspace{1.5cm}

\begin{center}
  {\bf Abstract}
\end{center}
In the weak coupling limit of $\su{N}$ Yang-Mills theory and the O($N$) vector model, explicit state counting allows us to
demonstrate the existence of a partially deconfined phase: 
$M$ of $N$ colors deconfine, and $\frac{M}{N}$ gradually grows from zero (confinement) to one (complete deconfinement). We point out that the mechanism admits a simple interpretation in the form of spontaneous breaking of gauge symmetry. 
In terms of the dual gravity theory, such breaking occurs during the formation of a black hole. We speculate whether
the breaking and restoration of gauge symmetry can serve as an alternative definition of the deconfinement transition
in theories without center symmetry, such as QCD. 
We also discuss the role of the color degrees of freedom in the emergence of the bulk geometry in holographic duality.

\newpage
\tableofcontents

\section{Introduction}
\hspace{0.51cm}
In this paper, we study the mechanism of the deconfinement transition in
large-$N$ gauge theory. 
Concretely, we establish the existence of the recently proposed {\it partially deconfined phase} \cite{Hanada:2016pwv,Hanada:2018zxn,Berenstein:2018lrm}, 
in simple, analytically tractable examples: The Gaussian matrix model, weakly coupled Yang-Mills theory on the three-sphere, 
and the free gauged vector model on the two-sphere. 

\begin{figure}[htbp]
\begin{center}
\scalebox{0.4}{
\includegraphics{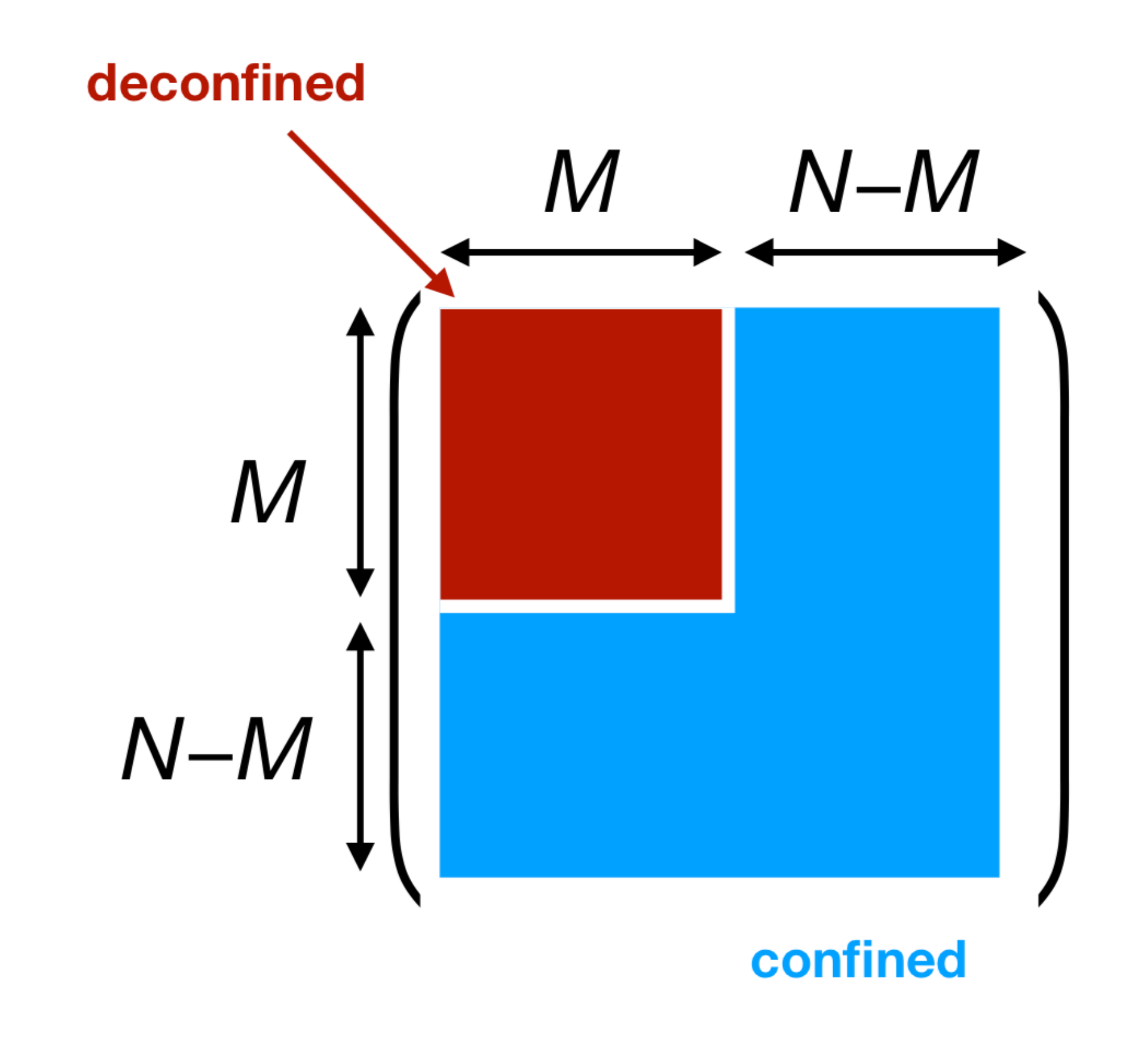}}
\end{center}
\caption{Partial deconfinement in the gauge sector and adjoint matters. 
Only the $M\times M$-block shown in red is excited. 
The consistency of this cartoon picture with the gauge singlet constraint is discussed around \eqref{eq:superposition}.
}\label{fig:matrix}
\end{figure}

Partial deconfinement means that a part of the $\su{N}$ gauge group, which we denote by $\su{M}$, deconfines, while  
the rest of the degrees of freedom are not excited; see Fig.~\ref{fig:matrix}.
It applies to other groups such as $O(N)$ as well.
Partial deconfinement has been introduced
in order to explain the thermal properties of 4d ${\cal N}=4$ super Yang-Mills on the three-sphere \cite{Hanada:2016pwv}.\footnote{
A similar idea with the same motivation, applied to different models, can be found in Refs.~\cite{Berkowitz:2016znt,Berkowitz:2016muc}
and Ref.~\cite{Asplund:2008xd}. 
}
According to the AdS/CFT duality \cite{Maldacena:1997re}, the thermodynamics of 4d ${\cal N}=4$ super Yang-Mills theory
is equivalent to that of the black hole in AdS$_5\times$S$^5$ \cite{Witten:1998zw}. 
Correspondingly, there must exist a phase dual to the small black hole which behaves as $E\sim N^2T^{-7}$ \cite{Aharony:1999ti}. 
Partial deconfinement naturally explains this behavior \cite{Hanada:2016pwv, Hanada:2018zxn}
due to the changing number of degrees of freedom participating in the dynamics \cite{Berkowitz:2016znt,Berkowitz:2016muc}. 

In Ref.~\cite{Hanada:2018zxn}, it has been conjectured that partial deconfinement should take place in any gauge theory
at sufficiently large $N$.\footnote{
This conjecture originates from the startling resemblance between the dynamics of D-branes 
and the phenomenological model of the formation of ant trails \cite{Beekman9703}. 
See Ref.~\cite{Hanada:2018zxn} for details. 
}  
It is natural to identify the Gross-Witten-Wadia (GWW) transition \cite{Gross:1980he,Wadia:2012fr},
which is characterized by the formation of a gap 
in the distribution of the Polyakov line phases, with the transition from the partially deconfined phase 
to the completely deconfined phase \cite{Hanada:2018zxn}.  
The deconfinement transition in the usual sense, where the distribution 
changes from uniform to nonuniform, 
corresponds to the Hagedorn transition \cite{Hagedorn:1965st}. 

The idea of partial deconfinement has withstood a number of consistency checks  \cite{Hanada:2016pwv,Hanada:2018zxn}, but no direct evidence, let alone an explicit construction, has been given so far. This gap will be filled in this paper.
We will show how in simple models, the partially deconfined states can be explicitly constructed in the Hilbert space. As we will see, the entropy is precisely explained by such states. 

Confinement in the theories we consider is due to a singlet constraint that is inherited from their interacting ancestors.
Despite their simple nature, the models exhibit a rich thermal structure \cite{Witten:1998zw, Sundborg:1999ue, Aharony:2003sx}. At low temperatures, their free energies are ${\cal O}(N^0)$ reminiscent of a completely confined phase where only singlet states can be excited. At high temperatures, on the other hand, their free energy becomes ${\cal O}(N)$ for fundamental matter, or ${\cal O}(N^2)$ for adjoint matter, implying that all individual degrees of freedom are excited. In this paper, we will demonstrate that the transition from low to high temperature phases goes through a phase of partial deconfinement.

We start with explaining basic properties of the partially deconfined phase in Sec.~\ref{sec:Partial_deconfinement_review}. 
Then in Sec.~\ref{sec:matrix_model}, Sec.~\ref{sec:4d_YM} and Sec.~\ref{sec:free-vector-model} 
we show partial deconfinement 
for the weak coupling limit of the matrix model, the weakly coupled Yang-Mills theory and the free gauged $O(N)$ vector model, respectively. 
Our arguments can be extended to interacting theories 
provided an assumption (`truncation to $M$ colors'), which will be explained explicitly, remains true. 
In Sec.~\ref{sec:gauge_symmetry_breaking}, we discuss in which sense the partially deconfined phase spontaneously breaks gauge symmetry. 
Sec.~\ref{sec:discussions} is devoted to the discussions.

\section{Basic properties of the partially deconfined phase}\label{sec:Partial_deconfinement_review}
\hspace{0.51cm}
We begin by introducing the basic concept of partial deconfinement at the hand of several defining properties. These have been used for a consistency check of the mechanism \cite{Hanada:2016pwv,Hanada:2018zxn} and 
will play important roles in this paper as well. 

For concreteness, we consider a confining $\su{N}$ gauge theory at large $N$, and an $\su{M}$ subgroup with $\frac{M}{N}$ of order $N^0$, s.t. we can ignore both $1/M$ and $1/N$ corrections. 
Let us further assume that 
the $\su{M}$ sector can be described well by ignoring interactions with the rest, as is the case for all theories studied in this paper.\footnote{ 
There are exceptions that require a modified set of criteria. We explain this in detail in Appendix~\ref{sec:super_renormalizable}. 
}
In this case, we can truncate the $N\times N$ matrices to $M\times M$ \cite{Hanada:2016pwv}, leading to an $\su{M}$ gauge theory that describes equivalent physics at low energies. 
Eventually, however, as we increase the energy, the $M\times M$ sector deconfines, and we need a larger subgroup to capture the physics of the full $\su{N}$ theory. This becomes clear when considering the free energy. In the $\su{M}$ sector it remains of order $M^2$, while in the full $\su{N}$ theory it grows towards $N^2$. The fact that the full $\su{N}$ theory is described by gradually growing completely deconfined subgroups is the essence of partial deconfinement. As we have stated before, it is natural to assume that complete deconfinement sets in at the critical point of the GWW transition \cite{Hanada:2018zxn}. 
Hence, the SU($M$)-deconfined sector of the SU($N$) theory should be seen as
the GWW-point of the SU($M$)-truncated theory \cite{Hanada:2018zxn}. 

It immediately follows that
$M$ of the Polyakov loop phases should follow the distribution at the GWW transition, 
while the other $N-M$ phases are distributed uniformly as in the confined phase. 
Namely, we expect the distribution of the phase $\theta$ ($-\pi\le\theta\le\pi$) to be \cite{Hanada:2018zxn}
\begin{eqnarray}
\rho(\theta)
&=&
\left(
1
-
\frac{M}{N}
\right)
\rho_{\rm confine}(\theta)
+
\frac{M}{N}
\cdot
\rho_{\rm GWW}(\theta; M)
\nonumber\\
&=&
 \frac{1}{2\pi}
\left(
1
-
\frac{M}{N}
\right)
+
\frac{M}{N}
\cdot
\rho_{\rm GWW}(\theta; M). 
\label{eq:Polyakov_loop}
\end{eqnarray}
All distributions above are normalized such that 
the integral from $-\pi$ to $+\pi$ is 1. 
Here, $\rho_\text{confine}(\theta)$ denotes the distribution in the confined phase, $\rho_\text{confine}(\theta) = \tfrac{1}{2\pi}$, while $\rho_{\rm GWW}(\theta; M)$ denotes the distribution at the GWW point of the $\su{M}$ theory. 
It can in principle depend on $M$, but in the examples we study in this paper, 
there is no $M$ dependence. 

Since $M \gg 1$, the entropy $S$ is dominated by the deconfined sector  
and, as we have explained above, it is natural to assume that the $\su{M}$ deconfined sector is approximated by the $\su{M}$ theory at the GWW transition. 
Therefore, one should expect \cite{Hanada:2016pwv,Hanada:2018zxn} 
\begin{eqnarray}
S=S_{\rm GWW}(M), 
\label{eq:entropy}
\end{eqnarray}
where the right hand side represents the entropy of the $\su{M}$ theory at the GWW transition. 
Other quantities, such as the energy, should behave in the same manner, namely 
\begin{eqnarray}
E=E_{\rm GWW}(M)
\label{eq:energy}
\end{eqnarray}
up to the zero-point energy \cite{Hanada:2016pwv,Hanada:2018zxn}. The fact that Eqs.\eqref{eq:Polyakov_loop}, \eqref{eq:entropy} and \eqref{eq:energy} are satisfied by a single $M$ serves as a strong consistency check. 

\section{Gauged Gaussian Matrix Model}\label{sec:matrix_model}
\hspace{0.51cm}
We begin with the simplest possible example, the gauged Gaussian matrix model. 
The Euclidean action is given by 
\begin{eqnarray}
S
=
N\sum_{I=1}^D\int_0^\beta dt {\rm Tr}
\left(
\frac{1}{2}(D_tX_I)^2
+
\frac{1}{2}X_I^2
\right),  
\end{eqnarray} 
where the number of matrices $D$ is larger than 1. 
The circumference of the temporal circle $\beta$ is related to temperature $T$ by $\beta=\frac{1}{T}$. 
The covariant derivative is defined by $D_tX_I=\partial_t X_I -i[A_t,X_I]$, 
where $A_t$ is the gauge field which is responsible for the gauge singlet constraint.

\begin{figure}[htbp]
\begin{center}
\scalebox{0.12}{
\includegraphics{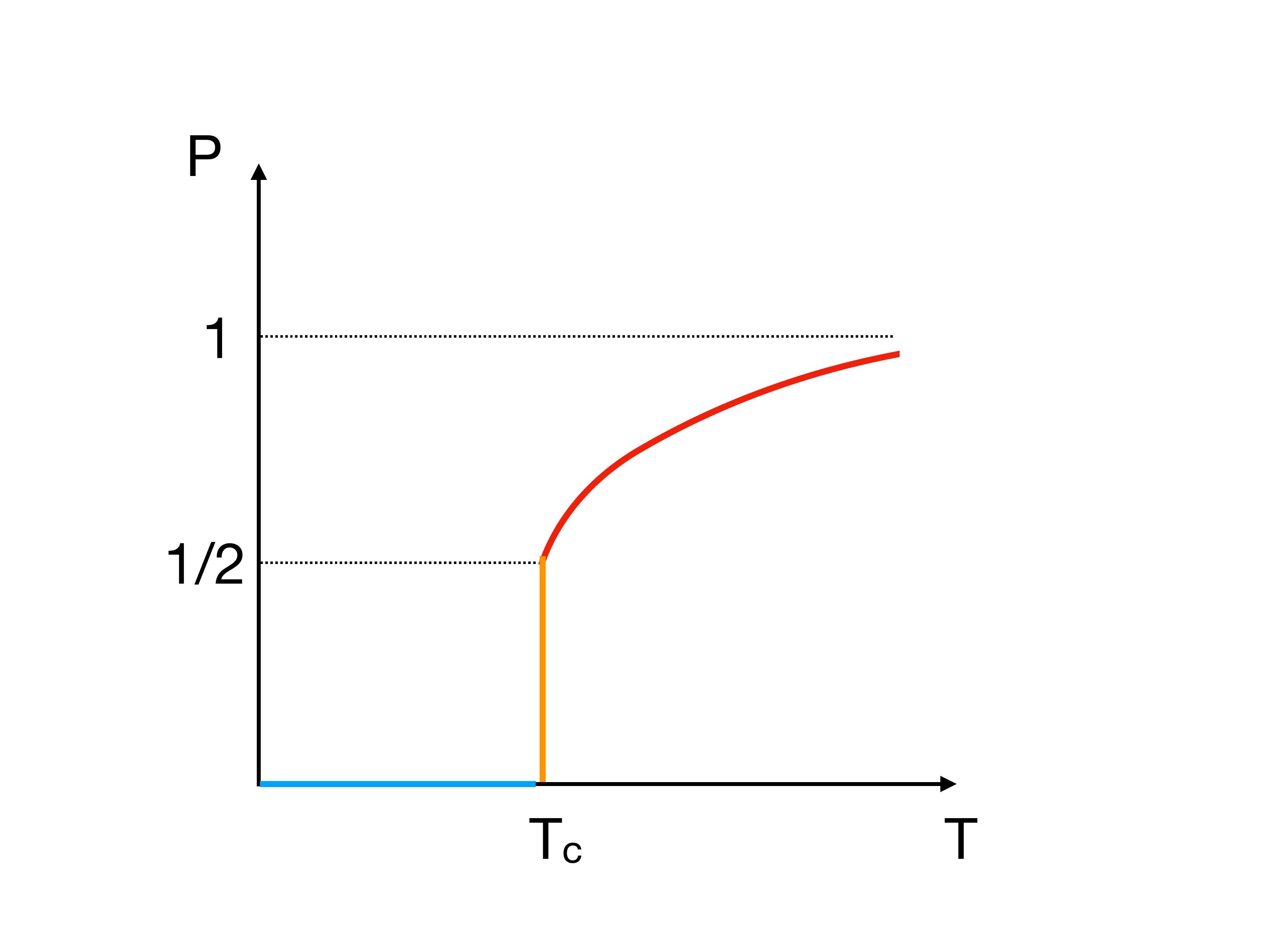}}
\scalebox{0.12}{
\includegraphics{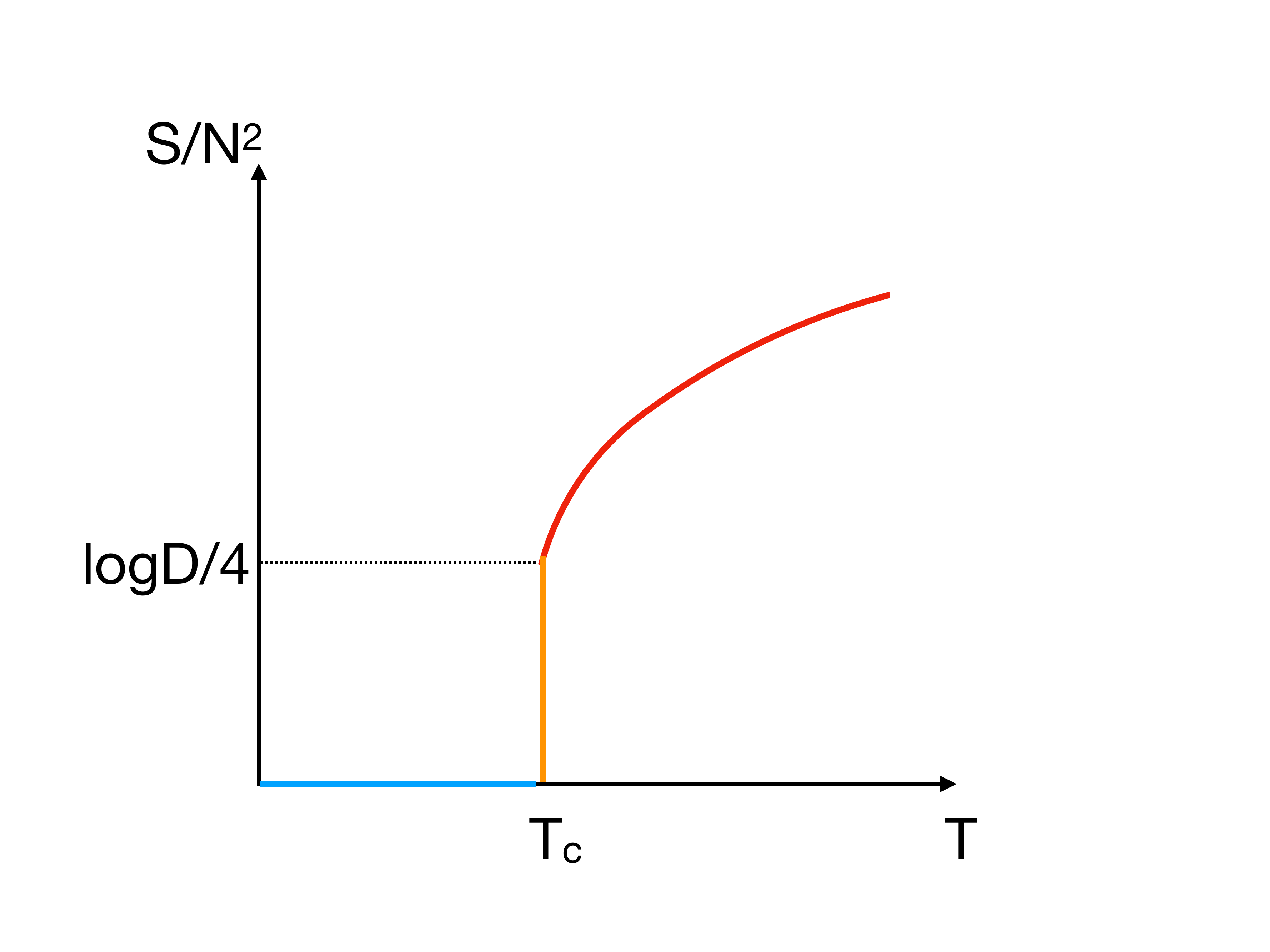}}
\scalebox{0.12}{
\includegraphics{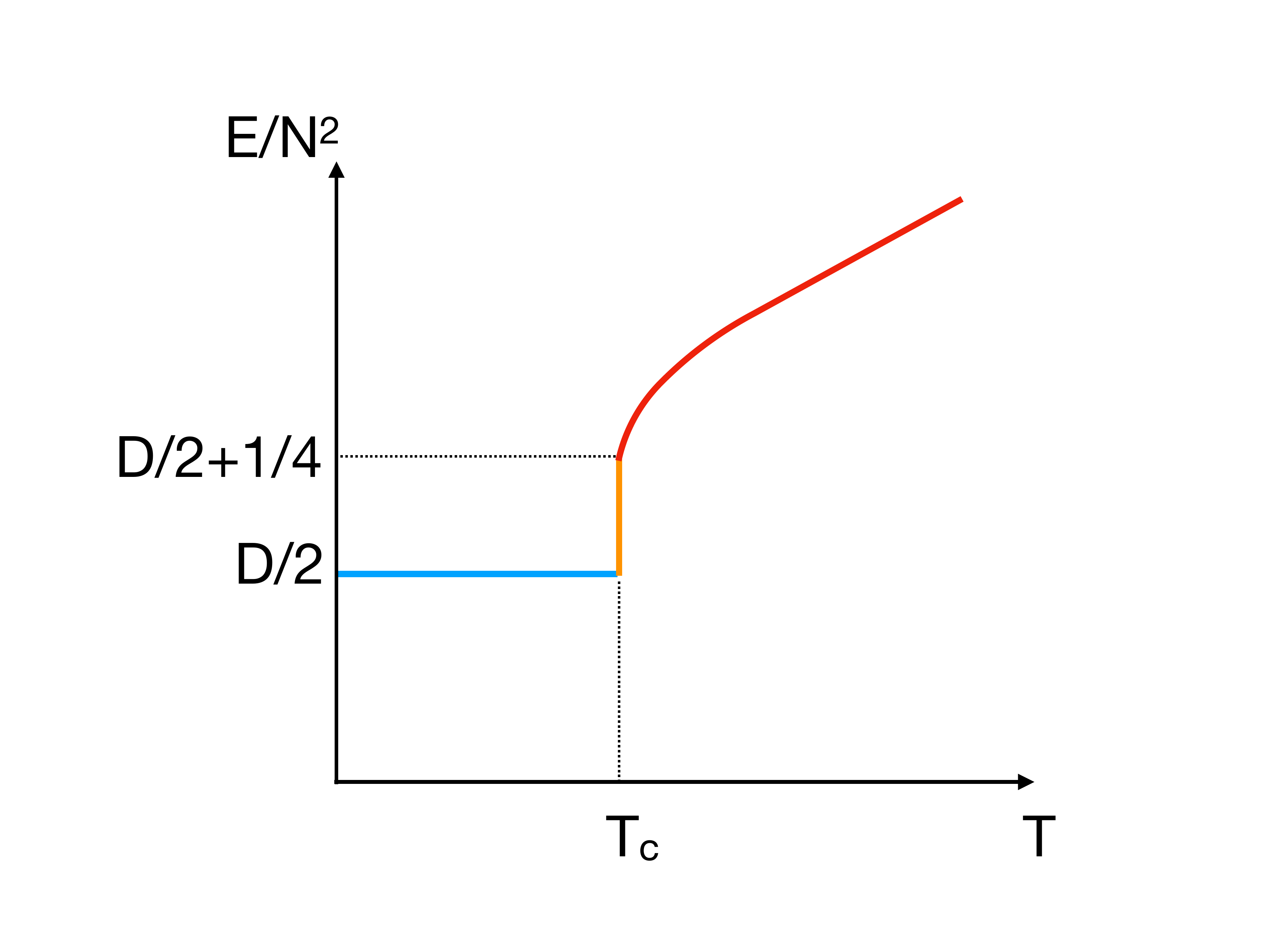}}
\end{center}
\caption{
Cartoon pictures of the temperature dependence of the Polyakov loop $P$, entropy $S$ and energy $E$ in the gauged Gaussian matrix model. 
Blue, orange and red lines are identified with the confined, partially deconfined and completely deconfined phases, respectively. 
}\label{fig:transition_GMM}
\end{figure}

The free energy  is given by
\begin{eqnarray}
\beta F 
&=&
-\log Z(\beta)
\nonumber\\
&=&
\frac{N^2D}{2}
\log\left(\det\left(-D_0^2+1\right)\right)
-
\frac{N^2}{2}
\log\left(\det\left(-D_0^2\right)\right)
\nonumber\\
&=&
\frac{DN^2\beta}{2}
+
N^2\sum_{n=1}^\infty
\frac{1-Dx^n}{n}|u_n|^2,
\end{eqnarray}
where $x=e^{-\beta}$, $u_n=\frac{1}{N}{\rm Tr}{\cal P}^n$, 
and ${\cal P}$ is the Polyakov line operator obtained from the holonomy of $A_t$ around the thermal circle. In the second line, 
$\frac{N^2D}{2}
\log\left(\det\left(-D_0^2+1\right)\right)$ is the contribution from $D$ scalars, 
while $\frac{N^2}{2}
\log\left(\det\left(-D_0^2\right)\right)$ is the Faddeev-Popov term corresponding to the static diagonal gauge. In the last line we have used \cite{Aharony:2003sx}
\begin{eqnarray}
\frac{1}{2}
\log\left(\det\left(-D_0^2+\omega^2\right)\right)
=
\frac{\beta\omega}{2}
-
\sum_{n=1}^\infty
\frac{x^{\omega n}}{n}|u_n|^2.
\end{eqnarray}
The $u_n$'s of a stable configuration of the model should minimize the free energy. 
At $T<T_c=\frac{1}{\log D}$, $|u_n|=0$ ($n\ge 1$) is favored. 
At $T=T_c$, $|u_1|$ can take any value from 0 to $\frac{1}{2}$, 
while $u_2,u_3,\cdots$ remain zero, without changing the free energy.  
This is a first order deconfinement transition.
In Fig.~\ref{fig:transition_GMM}, we have shown how the Polyakov loop $P$, entropy $S$ and energy $E$ depend on temperature.  

Let us consider the deconfinement transition at $T=T_c$. 
\begin{itemize}
\item
The Polyakov loop $P=|\frac{1}{N}{\rm Tr}{\cal P}|=|u_1|$ can take any value between 0 and $\frac{1}{2}$ at $T=T_c$.  
From the microcanonical viewpoint, $P=0, T=T_c$ is the transition point from confinement to deconfinement, 
and $P=\frac{1}{2}, T=T_c$ is the GWW transition. We will interpret them as the transitions from confinement to 
partial deconfinement, and from partial deconfinement to complete deconfinement.

\item
As derived in Appendix~\ref{sec:derivation_energy_free_YM}, at $T=T_c$, the energy can be written as 
\begin{eqnarray}
E(T=T_c, P,N)
=
\frac{DN^2}{2}
+
N^2P^2. 
\label{eq:energy_MM}
\end{eqnarray}
The first term is the zero-point energy. 

\item
The entropy $S=\beta(E-F)$ is 
\begin{eqnarray}
S(T=T_c, P,N)
=
N^2P^2\log D. 
\label{eq:entropy_MM}
\end{eqnarray}
\item
For any $N$, the distribution of the Polyakov line phases at the critical temperature is\footnote{
Strictly speaking, there is an ambiguity associated with the center symmetry, 
namely the constant shift of all the phases. We fixed the center symmetry such that the Polyakov loop becomes 
real and non-negative, i.e. $P=|P|$. 
} 
\begin{eqnarray}
\rho(\theta)=
\frac{1}{2\pi}\left(
1+2P\cos\theta\right)
=
\left(
1-2P
\right) 
\cdot
\frac{1}{2\pi}
+
2P\cdot
\frac{1}{2\pi}\left(
1+\cos\theta
\right),
\end{eqnarray}
up to $1/N$ corrections. In particular, the distribution at the GWW transition is 
\begin{eqnarray}
\rho_{\rm GWW}(\theta; N)
=
\frac{1}{2\pi}
\left(
1+\cos\theta
\right). 
\end{eqnarray}

\end{itemize}

These results are consistent with \eqref{eq:Polyakov_loop}, \eqref{eq:entropy} and \eqref{eq:energy}, 
with the identification of $\frac{M}{2N}=P$, $S_{\rm GWW}(M)=\frac{\log D}{4}M^2$ and  
$E_{\rm GWW}(M)=\frac{M^2}{4}$ up to the zero-point energy $\frac{DN^2}{2}$.  
Therefore, partial deconfinement is a promising description of this phase transition. 

Next we construct all the $\su{M}$ partially deconfined states in an $\su{N}$-invariant manner, 
and show that they precisely explain the entropy of the $\su{N}$ theory at $E=E_{\rm GWW}(M)$.\footnote{
Berenstein \cite{Berenstein:2018lrm} performed this task for the case of $D=2$.  
He pointed out that the states with $E\sim M^2$ can be represented by the Young tableaux 
with approximately $M$ rows and columns, and that they can be identified with the partially deconfined states. 
Below, we will give a more precise counting, including the $O(1)$ factors. 
Our method can easily be generalized to other theories. 
}
To this end, we will first, in an abstract manner, identify the orthonormal set of states in the $\su{M}$ theory that reproduce the entropy Eq.\eqref{eq:entropy_MM}.
Naturally, these are precisely the $\su{M}$ singlet states. 
Next, we extend them to orthonormal Hamiltonian eigenstates in the full $\su{N}$ singlet theory. To leading order in $1/M$ and $1/N$, these comprise the entire set of states that reproduces the correct scaling for the energy \eqref{eq:energy_MM} and entropy \eqref{eq:entropy_MM}. Remarkably, we will along the way encounter a notion according to which the $\su{N}$ symmetry is `spontaneously broken' to $\su{M} \times \su{N-M} \times {\rm U}(1)$.

Concretely, let us recall how the theory is defined in the Hamiltonian formulation. 
The Hamiltonian is given by 
\begin{eqnarray}
\hat{H}_{\rm free}
=
\sum_I
{\rm Tr}\left(
\frac{1}{2}\hat{P}_I^2
+
\frac{1}{2}\hat{X}_I^2
\right)
=
\sum_{I,\alpha}\left(
\frac{1}{2}\hat{P}_{I\alpha}^2
+
\frac{1}{2}\hat{X}_{I\alpha}^2
\right),  
\end{eqnarray} 
where $\hat{P}_{ij}=\sum_{\alpha=1}^{N^2-1}\hat{P}_\alpha\tau^\alpha_{ij}$,
$\hat{X}_{ij}=\sum_{\alpha=1}^{N^2-1}\hat{X}_\alpha\tau^\alpha_{ij}$, 
with properly normalized $\su{N}$ generators $\tau^\alpha$, 
and the commutation relation is given by $[\hat{X}_{I\alpha},\hat{P}_{J\beta}]=i\delta_{IJ}\delta_{\alpha\beta}$.  
Physical states are identified as those that satisfy the $\su{N}$-singlet condition.
Such singlet states can be constructed for example as ${\rm Tr}(\hat{a}^\dagger_I\hat{a}^\dagger_{I'}\cdots)|0\rangle$, 
where the ground state $|0\rangle$ is given as usual by $\hat{a}_{I\alpha}|0\rangle=0$. 

We now construct the $\su{M}$ subsector.
To this end, we separate the $\su{N}$ generators into an SU$(M)$ part $\tau^{\alpha'}$ (red block in Fig.~\ref{fig:matrix}) and 
the rest $\tau^{\alpha''}$ (blue block in Fig.~\ref{fig:matrix}). 
Similarly, we also separate the Hamiltonian into an SU$(M)$ part $\hat{H}_{\rm free}'$ and the rest $\hat{H}_{\rm free}''$. 
Explicitly, we have
\begin{eqnarray}
\hat{H}'_{\rm free}
=
\sum_{I,\alpha'}\left(
\frac{1}{2}\hat{P}_{I\alpha'}^2
+
\frac{1}{2}\hat{X}_{I\alpha'}^2
\right),  
\qquad
\hat{H}''_{\rm free}
=
\sum_{I,\alpha''}\left(
\frac{1}{2}\hat{P}_{I\alpha''}^2
+
\frac{1}{2}\hat{X}_{I\alpha''}^2
\right).   
\end{eqnarray} 
By construction, 
the density of states of an $\su{M}$-singlet theory correctly reproduces the energy and entropy \eqref{eq:energy_MM} and \eqref{eq:entropy_MM}. In the full $\su{N}$ Hilbert space, we construct the $\su{M}$-invariant states by acting on the vacuum only with $\alpha'$-operators $\hat{a}^\dagger_{I\alpha'}$. Let us refer to them as $|E;{\rm SU}(M)\rangle$. 
Due to the absence of the interaction term, 
they are eigenstates of the full Hamiltonian $\hat{H}_{\rm free}=\hat{H}'_{\rm free}+\hat{H}''_{\rm free}$,
with\footnote{This is the only place where we use the assumption that the interaction is absent.}
\begin{eqnarray}
\hat{H}'_{\rm free}|E;{\rm SU}(M)\rangle=E|E;{\rm SU}(M)\rangle. 
\end{eqnarray}
and
\begin{eqnarray}
\hat{H}''_{\rm free}|E;{\rm SU}(M)\rangle
=
\frac{D(N^2-M^2)}{2}|E;{\rm SU}(M)\rangle. 
\end{eqnarray}

In order to project $|E;{\rm SU}(M)\rangle$ into the $\su{N}$ singlet sector, we 
consider the superposition of such states over all possible $\su{N}$ gauge transformations ${\cal U}$ that rotate the ladder operators $\hat{a}$, 
\begin{eqnarray}\label{Einv}
|E\rangle_{\rm inv}
\equiv
{\cal N}^{-1/2}\int dU\,{\cal U}\left(|E;{\rm SU}(M)\rangle\right), 
\label{eq:superposition}
\end{eqnarray}
where the integral runs over $\su{N}$ by using the Haar measure, and ${\cal N}$ is a normalization factor. It is not difficult to see that $|E\rangle_{\rm inv}$ is not zero, and that $|E\rangle_{\rm inv}$'s made of different $\su{M}$-invariant states are linearly independent. We refer the interested reader to Sec.~\ref{sec:nonzero-norm} for a detailed discussion. 
Since by construction, $|E\rangle_{\rm inv}$ is still an eigenstate of $\hat{H}_{\rm free}$ with the same energy, 
we have constructed an explicit mapping between the eigenstates of $\su{M}$ and SU$(N)$ theories in a gauge-invariant manner, 
without changing the energy. 
By using the states $|E=E_{\rm GWW}(M)\rangle_{\rm inv}$, 
the entropy of the $\su{N}$ theory is explained, and hence, no other states are needed. 

The fact that $M$ eigenvalues of the constraining matrix are deconfined, while $N-M$ are still confined,
points to an effective breaking of the $\su{N}$ symmetry down to an $\su{M} \times \su{N-M} \times {\rm U}(1)$ subgroup. 
We will discuss this point in detail in Sec.~\ref{sec:gauge_symmetry_breaking}.

\subsection*{Why ${\rm SU}(M)\times{\rm SU}(N - M)\times {\rm U}(1)$?}
\hspace{0.51cm}
If we just demanded the entropy to be $S \sim M^2$ for a given energy, deconfining a
${\rm SU}(M_1)\times{\rm SU}(M_2)\times\cdots$-sector with $M_1^2+M_2^2+\cdots=M^2$
would appear equally valid.   
However in order to explain the value of the Polyakov loop, we are bound to consider exactly the $\su{M}$-deconfined phase.
In order to understand why this is the case, note that we have introduced a small interaction term in order to derive 
\eqref{eq:energy_MM} and \eqref{eq:entropy_MM} (see Appendix~\ref{sec:derivation_energy_free_YM}). With this interaction, 
Polyakov line phases typically attract each other and prefer a single bound state. Therefore, SU$(M)$-deconfinement minimizes the free energy, as opposed to all other patterns.  
The same comment applies to our next example explained in Sec.~\ref{sec:4d_YM}. 

\section{Weakly coupled Yang-Mills theory on S$^3$}\label{sec:4d_YM}
\hspace{0.51cm}
Another theory in which the mechanism of partial deconfinement is explicitly tractable is given by the free limit of Yang-Mills on S$^3$. 
This theory can be solved analytically,
and captures important features of the deconfinement transition \cite{Sundborg:1999ue,Aharony:2003sx}. 

Due to the curvature of the spatial S$^3$, all the modes except for the Polyakov line phases become massive and can be integrated out to construct an effective action for the Polyakov line phases.
The results relevant for us are as follows \cite{Sundborg:1999ue,Aharony:2003sx}. 
\begin{itemize}
\item
There is a first order deconfinement phase transition at $T=T_c$. 
The Polyakov loop $P$ can take any value between 0 and $\frac{1}{2}$ at the critical temperature.  
From the microcanonical viewpoint, $P=0, T=T_c$ is the transition point from the confinement to deconfinement, 
and $P=\frac{1}{2}, T=T_c$ is the GWW transition. We will interpret them as the transition from confinement to 
partial deconfinement, and from partial deconfinement to complete deconfinement.

\item
At the critical temperature, the energy can be written as\footnote{
As before, we are considering an infinitesimally small interaction, rather than considering literally `free' theory.  
} 
\begin{eqnarray}
E(T=T_c, P,N)
=
E_{\rm GWW}(N)\times |2P|^2
\propto N^2 \times |2P|^2, 
\end{eqnarray}
up to the zero-point energy which is proportional to $N^2$. 
The entropy is 
\begin{eqnarray}
S(T=T_c, P,N)
=
S_{\rm GWW}(N)\times |2P|^2
\propto N^2 \times |2P|^2.  
\end{eqnarray}

\item
For any large 
 $N$, the distribution of the Polyakov line phases at the critical temperature is 
\begin{eqnarray}
\rho(T=T_c, \theta; N)=\frac{1}{2\pi}\left(1+2P\cos\theta\right)
\end{eqnarray}
up to $1/N$ corrections. In particular, the distribution at the GWW transition is 
\begin{eqnarray}
\rho_{\rm GWW}(\theta; N)=\frac{1}{2\pi}\left(1+\cos\theta\right).\  
\end{eqnarray}

\end{itemize}

These results are consistent with \eqref{eq:Polyakov_loop}, \eqref{eq:entropy} and \eqref{eq:energy}, 
with the identification of $P=\frac{M}{2N}$. 
Therefore, partial deconfinement is a promising description of this phase transition. 

In order to construct the partially deconfined states explicitly, 
we can repeat the argument presented in Sec.~\ref{sec:matrix_model}. 
Namely we construct all $\su{M}$ partially deconfined states in an $\su{N}$-invariant manner, 
and show that the entropy of the $\su{N}$ theory at $E=E_{\rm GWW}(M)$ is explained precisely by those states. 
Let us use $\hat{\phi}_\alpha$ and $\hat{\pi}_\alpha$ to denote the fields and the conjugate momenta. 
As before, we split them into two sets: the $\su{M}$ subsector ($\hat{\phi}_{\alpha'}$, $\hat{\pi}_{\alpha'}$), 
and the rest ($\hat{\phi}_{\alpha''}$, $\hat{\pi}_{\alpha''}$). 
We start with the eigenstates of the truncated SU$(M)$ theory, which are obtained by acting only with $\hat{\phi}_{\alpha'}$ and $\hat{\pi}_{\alpha'}$
on the $\su{N}$ invariant perturbative vacuum. At the transition point to deconfinement for the $\su{M}$ theory, there are by construction precisely enough such states with energy $E_{\rm GWW}(M)$ 
to explain the entropy $S(E=E_{\rm GWW}(M))=S_{\rm GWW}(M)$. 

In line with Sec.~\ref{sec:matrix_model}, 
we can trivially uplift the states to eigenstates of the SU$(N)$ theory with the same energy, up to zero-point contributions. 
Moreover, integrating over all gauge transformations on the three-sphere will project onto the $\su{N}$ singlet sector. Explicitly, 
due to the gauge invariance of the Hamiltonian, 
$|E\rangle_{\rm inv}\equiv
{\cal N}^{-1/2}
\int [dU] {\cal U}\left(|E;{\rm SU}(M)\rangle\right)$ is an
$\su{N}$-invariant eigenstate of $\hat{H}_{\rm free}$ when the integral is taken over the gauge transformation on the three-sphere. 

Thus, we have shown that there are just enough number of states in the `$\su{M}$-deconfined' sector. Correspondingly, no further macroscopic contributions can arise from other sectors, since this would contradict the analytic results above.

The argument in this section did not specify the regularization. 
For those who have concern about this point, we explain the lattice regularization in Appendix~\ref{sec:lattice_formulation}. 

Again, we are tempted to interpret that the gauge symmetry is broken as 
${\rm SU}(N)\to {\rm SU}(M)\times {\rm SU}(N-M)\times {\rm U}(1)$. 
In Sec.~\ref{sec:gauge_symmetry_breaking}, we will discuss this point further. 

\section{Free O($N$) Vector Model}\label{sec:free-vector-model}
\hspace{0.51cm}
We may also interpret the results obtained by Shenker and Yin \cite{Shenker:2011zf}
from the point of view of partial deconfinement, which provides yet another nontrivial consistency check.
We consider the 3d free O$(N)$ vector model on the two-sphere of unit radius
in the O($N$) singlet sector. For simplicity we set the number of flavor $N_f$ to be one. 
We can repeat essentially the same argument for generic dimensions $d$ and $N_f>1$. 
The minimal way to enforce the singlet constraint is through a Lagrange multiplier field. The deconfinement transition can then be studied by considering the effective action for the Lagrange multiplier after integrating out all massive excitations.
In $d=3$, this is equivalent to 
introducing a gauge field $A_\mu$ with Chern-Simons action, and taking the zero-coupling limit, and thus allows a description in forms of Polyakov line phases.

Compared to the matrix model, the biggest difference is the absence of Hagedorn-behavior in the vector model. 
Therefore, deconfinement takes place gradually as the energy increases. 
Indeed, the Polyakov loop is zero at $T=0$, nonzero at any $T>0$, 
and the Gross-Witten-Wadia transition, which is identified with the transition to complete deconfinement, 
takes place at $T=\frac{\sqrt{3}}{\pi}\sqrt{N}$. 

By using $b=\frac{T}{\sqrt{N}}$, and taking $b$ to be of order one, 
the distribution of the Polyakov line phase $\theta$ is written as\footnote{
Generalization to $N_f>1$ can be obtained by changing the definition of $b$ with $ b=T\sqrt{\frac{N_f}{N}}$,
as long as $\frac{N_f}{N}\ll 1$. 
}
\begin{eqnarray}
\rho(\theta)
=
\frac{1}{2\pi}
+
\frac{2b^2}{\pi}f(\theta), 
\end{eqnarray}
where 
\begin{eqnarray}
f(\theta)
=
-\frac{\pi^2}{12}+\frac{\left(|\theta|-\pi\right)^2}{4}.  
\end{eqnarray}
At $b=b_{\rm GWW}=\frac{\sqrt{3}}{\pi}$, the GWW transition takes place; the distribution becomes zero at $\theta=\pm\pi$.

\begin{figure}[htbp]
\begin{center}
\scalebox{0.4}{
\includegraphics{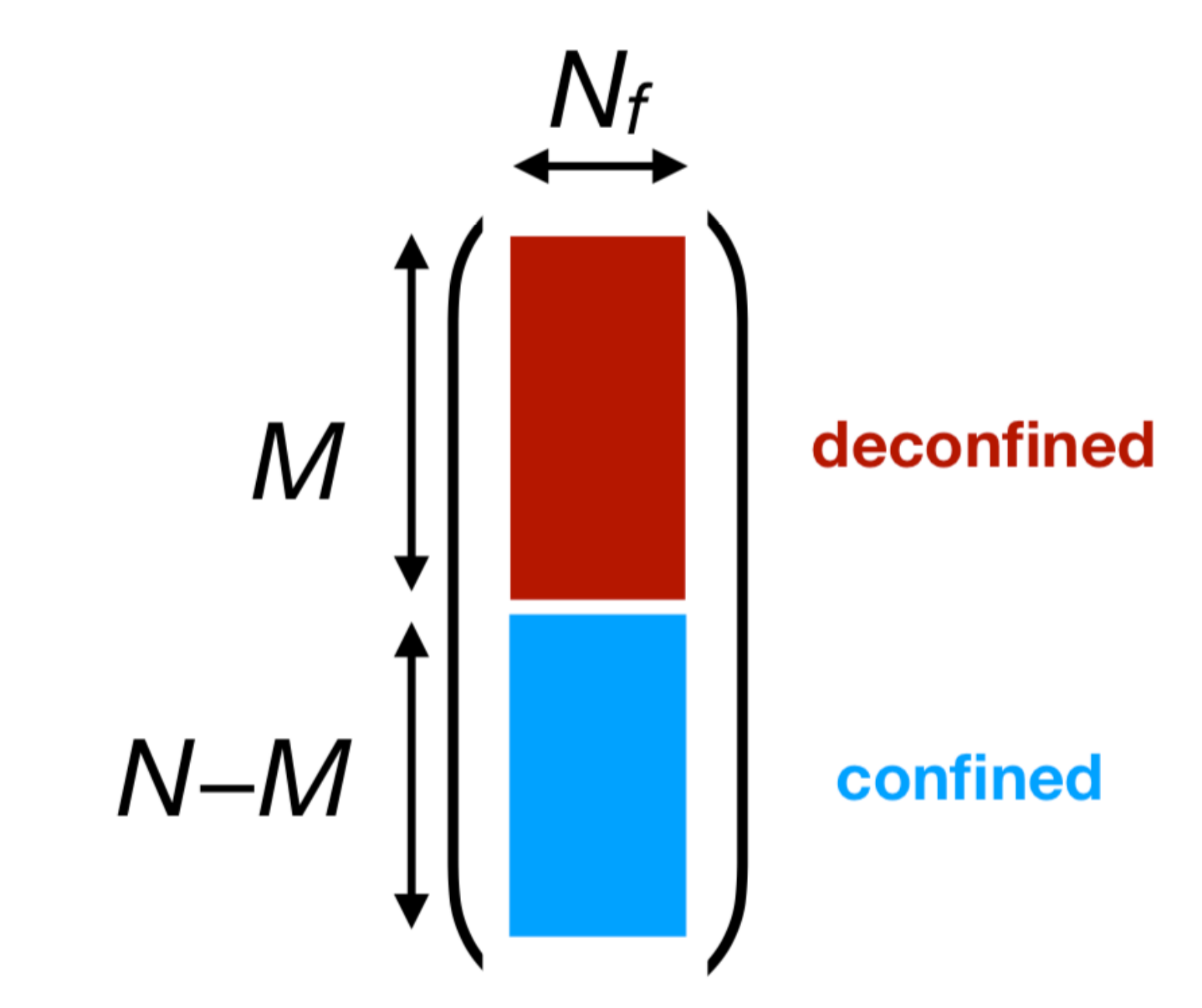}}
\end{center}
\caption{Partial deconfinement in the vector model. 
}\label{fig:vector}
\end{figure}

We can rewrite $\rho(\theta)$ as 
\begin{eqnarray}
\rho(\theta)
&=&
\frac{1}{2\pi}
-
\frac{\pi b^2}{6}
+
\frac{b^2\left(|\theta|-\pi\right)^2}{2\pi} 
\nonumber\\
&=&
\frac{1}{2\pi}
\left(
1
-
\frac{b^2}{b_{\rm GWW}^2}
\right)
+
\frac{b^2}{b_{\rm GWW}^2}
\cdot
\frac{b_{\rm GWW}^2\left(|\theta|-\pi\right)^2}{2\pi} 
\nonumber\\
&=&
\frac{1}{2\pi}
\left(
1
-
\frac{b^2}{b_{\rm GWW}^2}
\right)
+
\frac{b^2}{b_{\rm GWW}^2}
\cdot
\rho(\theta; b=b_{\rm GWW}). 
\end{eqnarray}
This relation, combined with \eqref{eq:Polyakov_loop}, suggests 
\begin{eqnarray}
\frac{M}{N}
=
\frac{b^2}{b_{\rm GWW}^2}, 
\label{eq:T-and-M-1}
\end{eqnarray}
where $M$ is the size of the deconfined sector (see Fig.~\ref{fig:vector}). Equivalently, 
\begin{eqnarray}
T
=
b\sqrt{N}
=
b_{\rm GWW}\sqrt{M}. 
\label{eq:T-and-M-2}
\end{eqnarray}
Note that the critical temperature of the `truncated' O($M$) theory is $T_{\rm GWW}(M)=b_{\rm GWW}\sqrt{M}$. 
Therefore, the identification leads to 
\begin{eqnarray}
\rho(\theta,T=T_{\rm GWW}(M))
=
 \frac{1}{2\pi}
\left(
1
-
\frac{M}{N}
\right)
+
\frac{M}{N}
\cdot
\rho_{\rm GWW}(\theta; M). 
\end{eqnarray}

At $1 \ll T\le T_{\rm GWW}$, the energy scales as
\begin{eqnarray}
E
=
A T^5, 
\end{eqnarray}
with an $N$-independent coefficient $A=16\,\zeta(5)$. 
Therefore, with $T=T_{\rm GWW}(M)$, the relation \eqref{eq:energy} holds, as well as relation \eqref{eq:entropy} with $S=\frac{5}{4}AT^4$. 
In this way, the Polyakov loop, energy and entropy are consistently explained by the same $M$
defined by \eqref{eq:T-and-M-1} and \eqref{eq:T-and-M-2}.  
Notice that unlike the previous cases with matrix degrees of freedom, for the free vector model the critical energy scales with $M$ as $E_{\rm GWW}(M)\sim M^{\frac{5}{2}}$.

We can easily repeat the argument presented in Sec.~\ref{sec:matrix_model} and Sec.~\ref{sec:4d_YM} to prove partial deconfinement.
The only difference is that the fields transform in the fundamental representation. 

Combining the results of this section with those of Sec.~\ref{sec:4d_YM} allows us to
perform a very similar analysis to the weak coupling limit of QCD on S$^3$ \cite{Schnitzer:2004qt}, corresponding to
weakly coupled Yang-Mills plus quarks in the fundamental representation. Again, we can  
confirm the existence of a partially deconfined phase. 
It is interesting to note that partial deconfinement appears to work without the $\mathbb{Z}_N$ center symmetry. 

\section{`Spontaneous breaking' of gauge symmetry}\label{sec:gauge_symmetry_breaking}
\hspace{0.51cm}
A potentially uncomfortable, but at the same time rather intriguing aspect of partial deconfinement is the apparent breaking of the $\su{N}$ gauge symmetry.\footnote{Here and in the following we will not be very picky in our choice of words and refer to the breaking of the global subgroup as ``gauge symmetry breaking''. We refer the reader to \cite{Elitzur:1975im}, where such subtleties are beautifully addressed.}
The aim of this section is to discuss this point and its potential applications to real world systems. Naturally, in particular the latter implies that some of our points will be rather speculative. 
We will frame our discussion in the language introduced in Sec. \ref{sec:matrix_model}. 

To recall, we have introduced states $|E;{\rm SU}(M)\rangle$ that alone account for the density of states at energy $E$. In order to make the state counting precise, out of $|E;{\rm SU}(M)\rangle$ we have constructed $\su{N}$ invariant states $|E\rangle_{\rm inv}$ in a one-to-one manner. So how does this go in hand with the notion of symmetry breaking?

To see this clearly, let us take a step back and not gauge the $\su{N}$ symmetry. In that case, the partition sum runs over states with arbitrarily charged states, and can in principle allow for saddles with spontaneously broken $\su{N}$ symmetry.
However, such saddles will lead to a gross overestimate of the density of states, 
since we are required to count all linearly independent states that can be obtained from $\su{N}$ rotating $|E;{\rm SU}(M)\rangle$ for arbitrary ${\cal U}$ as genuinely different entities if the symmetry is global.
 Luckily, there is a straightforward interpretation of this overestimate. By construction, only the transformations in the broken sector $\su{N}/[\su{M}\times\su{N-M}\times U(1)]$ act nontrivially on $|E;{\rm SU}(M)\rangle$. But these are of course nothing but the Nambu--Goldstone modes associated with the broken symmetries. And here gauge invariance comes to our rescue.

Just like in the Higgs mechanism, the Nambu--Goldstone modes are eliminated from the spectrum through gauge invariance. There, they are eaten by gauge bosons and become massive. In our case, the resolution is much more mundane, 
since the details of the spontaneous symmetry breaking is different from the conventional Higgs mechanism - we simply declare them to correspond to redundant transformations that do not change the physical state.

Thus, while through this trick the $\su{N}$ singlet nature of all states is restored, we may nevertheless use the picture of spontaneously broken gauge symmetry as a `convenient fiction' \cite{Rajagopal:2000wf}.
As such, we may attempt to extend it to more general setups with dynamical gluons, like those considered in the latter sections of the paper. Here, it proves convenient to 
fix part of the gauge symmetry, leaving only $\su{M}\times\su{N-M}\times$U(1). 
As long as we consider only SU($M$)-invariant operators in the deconfined block
(resp. the SU($N-M$)-invariant operators in the confined block), the system looks totally deconfined
(resp. confined). 
Off-diagonal blocks of the gauge field, which transform as a bifundamental field,
are not thermally excited, which suggests that they are massive.  
Probe quarks in the SU($M$) sector are deconfined (i.e. the quark-antiquark pair can be separated without major energy cost), while those in the SU($N-M$) sector are confined. 
The spectrum is clearly different compared to the `gauge symmetric phases', which are the completely confined and completely deconfined phases.
The gauge fixing makes the physics more easily accessible and `gauge symmetry breaking'
serves as a convenient fiction. 

In Refs.~\cite{Berkowitz:2016znt,Berkowitz:2016muc}, 
the D0-brane matrix model has been studied. This model has scalars, 
and the eigenvalues of scalars can escape to infinity along the flat direction.
It has been pointed out that the chain of the gauge symmetry breaking due to Higgsing, 
which is associated with the flat direction, naturally leads 
to negative specific heat.
The situation under consideration is similar to this case, from the point of view of string theory:
not all D-branes are in a bound state.
Similar gauge symmetry breaking by Higgsing is the key ingredient of the Matrix Theory proposal \cite{Banks:1996vh} which admits multi-body interactions to emerge from matrix degrees of freedom.
For theories without flat directions, such as 4d ${\cal N}=4$ SYM on S$^3$, we can consider essentially the same situations by using multiple deconfined blocks. 
This is reminiscent of the indistinguishability between the Higgsing and confinement \cite{Fradkin:1978dv}.
Again, gauge symmetry breaking provides us with a convenient fiction, making physics intuitively understandable.

\section{Discussions}\label{sec:discussions}
\hspace{0.51cm}
Once interactions are included, the $\su{M}$-invariant states are no longer exact energy eigenstates. 
However, because partial deconfinement appears to be a good picture in various interacting theories \cite{Hanada:2016pwv,Hanada:2018zxn},  
we find it likely that such an approximation is well-founded and the rest of the arguments in this paper can be applied without major change. 

\begin{figure}[htbp]
\begin{center}
\scalebox{0.4}{
\includegraphics{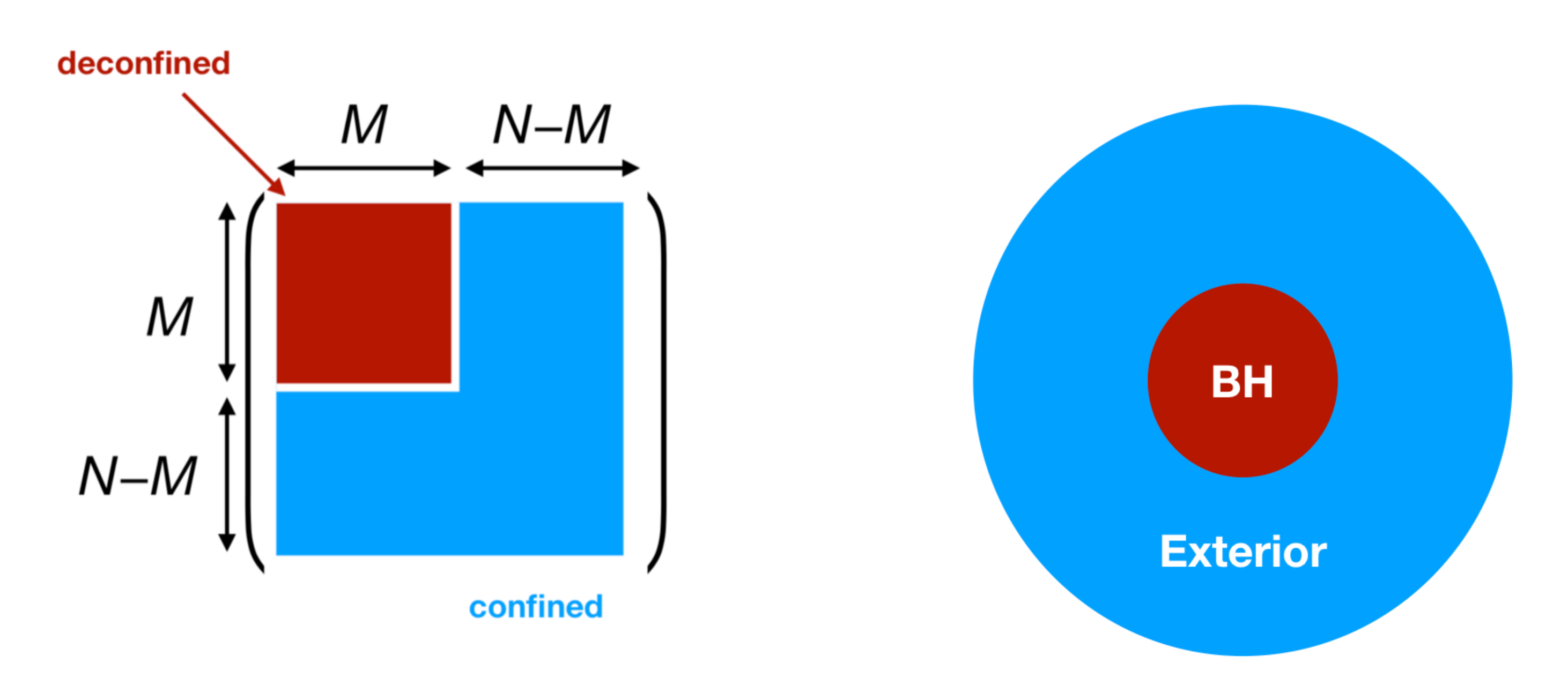}}
\end{center}
\caption{
A natural dual gravity interpretation of partial deconfinement. 
}\label{fig:BH-geometry}
\end{figure}

\begin{figure}[htb]
\begin{center}
\scalebox{0.4}{
\includegraphics{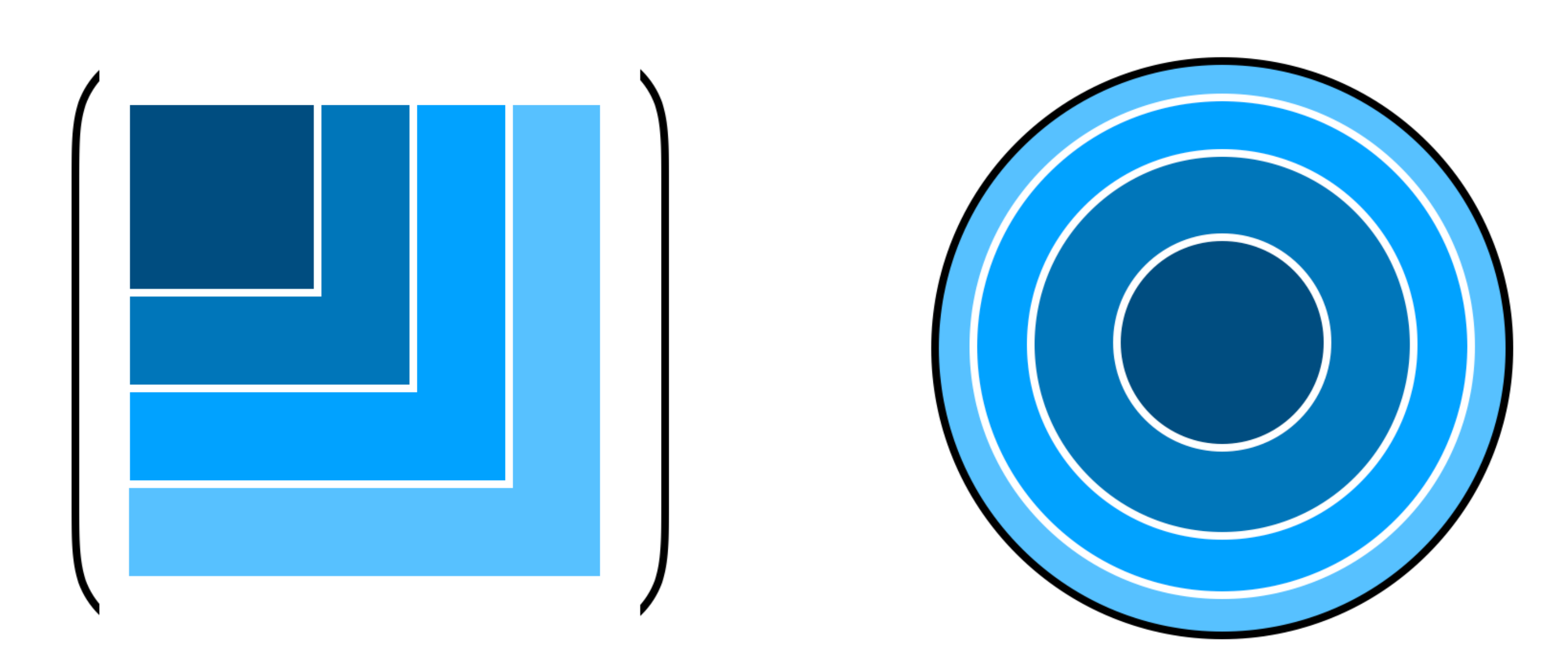}}
\end{center}
\caption{
A possible correspondence between the matrix degrees of freedom (left) and bulk geometry (right) suggested by partial deconfinement. 
As the black hole becomes larger, 
some part of the bulk geometry is hidden behind the horizon, and some part of the matrices deconfines, making it 
natural to identify them.   
Moreover, integrating out $M^2$ color degrees of freedom in the ground state (completely confining phase), 
naturally leads to an entanglement entropy of order $M^2$, given the all-to-all interaction, when $\frac{M}{N}\ll 1$. 
This is the same order as the entropy of the SU($M$)-deconfined phase, 
which is the area of the horizon of the small black hole. This is reminiscent of the observation that 
the entanglement entropy of two spatial regions is proportional to the area of the boundary between them \cite{Srednicki:1993im}.  
}\label{fig:bulk-vs-matrix}
\end{figure}

When a weakly-curved gravity dual is available, 
the most natural geometric interpretation of the partially deconfined phase would be that 
the deconfined and confined sectors describe the interior (or the horizon) and exterior of the black hole, respectively (Fig.~\ref{fig:BH-geometry}).\footnote{
Some discussions related to this issue can be found in Refs.~\cite{Hanada:2016pwv,Rinaldi:2017mjl}. 
} 
Because the confined sector is the same as the ground state up to the $1/N$-suppressed effects, 
the quantum entanglement expected in the ground state should survive, 
while in the deconfined sector the thermal excitations can break the entanglement. 
We are tempted to speculate that the strong entanglement in the confined sector is responsible for the emergence 
of the bulk geometry from the matrix degrees of freedom, along the line suggested by Van Raamsdonk \cite{VanRaamsdonk:2010pw}.  
Also, this picture suggests that the bulk geometry can be encoded in the matrices as shown in Fig.~\ref{fig:bulk-vs-matrix}, 
by slicing both sides to the layers in a natural manner and relating the row/column number and the radial coordinate.\footnote{
It would be interesting to speculate a possible connection bewteen the large-$N$ renormalization \cite{Brezin:1992yc}
and the holographic renormalization (see Ref.~\cite{Skenderis:2002wp} for a review). 
}$^,$\footnote{
The large-$N$ renormalization has been studied for the non-singlet sector of the $c=1$ matrix model, 
and the existence of the black hole phase with negative specific heat has been reported \cite{Dasgupta:2003xn}.
In such context, it would be interesting to generalize our argument to the non-singlet sector. 
}

Extrapolating our results to small $N$ and non-vanishing coupling,
a natural place to look for partial deconfinement and spontaneous gauge symmetry breaking is the physics of heavy-ion collisions. 
At low density, the thermal `transition' appears to be a rapid crossover \cite{Aoki:2006we}, 
which mimics a thermodynamically stable partially deconfined phase \cite{Hanada:2018zxn}, like for the models discussed in Sec.~\ref{sec:free-vector-model}. 
Therefore, in the cross-over region, partial deconfinement might be an approximately good description, which by the logic of Sec.~\ref{sec:free-vector-model} implies that gauge symmetry will be broken spontaneously. 
Close to the critical point, the deconfined sector would behave like the Hagedorn string. 
Another interesting possible consequence is the enhancement of the flavor symmetry, when an SU($2$) subgroup of the
SU($3$) gauge group is confined or deconfined: 
because SU($2$) is pseudo-real, the flavor symmetry is enhanced to SU($2N_f$) \cite{Coleman:1980mx,Peskin:1980gc}. 
Such enhancement of symmetry can change the spectrum drastically. 
The observations regarding the low-energy behavior of the Quark-Gluon-Plasma in Refs.~\cite{Rohrhofer:2019qwq,Denissenya:2014poa,Alexandru:2019gdm}
might be related to these scenarios.\footnote{
See Refs.~\cite{Shuryak:2019fgq,Glozman:2019mkx} for a recent debate regarding Refs.~\cite{Rohrhofer:2019qwq,Denissenya:2014poa}. 
} 

In QCD, the center symmetry does not exist because of the quarks in the fundamental representation. 
However, the arguments for partial deconfinement apply in the same manner, 
as mentioned in Sec.~\ref{sec:free-vector-model}. An intuitive paraphrase for the mechanism is that {\it if the energy is not large enough to excite all degrees of freedom, 
only a part of the degrees of freedom is excited}. 
Such a symmetry breaking mechanism would apply to other symmetries as well, whether they are gauged or not. 
As an example, let us consider 4d ${\cal N}=4$ SYM on S$^3$. 
The dual gravity description \cite{Witten:1998zw} naturally leads to two kinds of small black hole solutions with negative specific heat: the AdS$_5$-Schwarzschild solution which is not localized on S$^5$, 
and the `ten-dimensional' black hole which is localized on S$^5$ \cite{Aharony:1999ti}. The latter breaks the SO$(6)$ R-symmetry spontaneously. 
Possibly,
this phase diagram can be explained on the QFT side both by partial deconfinement and another closely related mechanism. While both solutions are partially deconfined in terms of the color degrees of freedom, the spontaneous breaking of the flavor symmetry would be due to an enhanced excitation of one of the scalars $X_1,X_2,\cdots,X_6$.
Such a scenario would allow us to resolve the puzzling features of the transition 
pointed out in Ref.~\cite{Yaffe:2017axl}.

In theories without center symmetry, such as QCD, it is difficult to precisely define `deconfinement'. 
In the large-$N$ limit, the jump of the energy from order one to order $N^2$ can serve this purpose. 
If one can extrapolate our argument to $N=3$, the breaking and restoration of gauge symmetry can give a good definition for deconfinement in real QCD.

\vspace{-0.5em}
\begin{center}
\subsection*{Acknowledgements}
\end{center}
\hspace{0.51cm}
We thank G.~Bergner, N.~Bodendorfer, 
A.~Cherman, G.~Dvali, N.~Evans, H.~Fukaya, 
C.~Gomez, S.~Hashimoto, G.~Ishiki, 
S.~Katz, A.~O'Bannon, R.~Pisarski, 
E.~Rinaldi, B.~Robinson, P.~Romatschke, A.~Sch\"{a}fer, S.~Shenker, 
H.~Shimada, K.~Skenderis, B.~Sundborg, M.~Tezuka, and H.~Watanabe for useful discussions and comments. 
MH thanks the Niels Bohr Institute and Brown University for the hospitality during his visits. CP thanks Shanghai Jiaotong University, Tianjin University and Jilin University for warm hospitality during the preparation of this paper. 
This  work  was partially supported by the STFC Ernest Rutherford Grant ST/R003599/1
and JSPS  KAKENHI  Grants17K1428. AJ and CP were supported by the US Department of Energy under contract DE-SC0010010 Task A. CP was also supported by the U.S. Department of Energy grant DE-SC0019480 under the HEP-QIS QuantISED program and by funds from the University of California. NW acknowledges support by FNU grant number
DFF-6108-00340.
\newpage

\appendix

\section{Some technicalities associated with the gauged Gaussian matrix model}
\hspace{0.51cm}
\subsection{Energy and entropy}\label{sec:derivation_energy_free_YM}
\hspace{0.51cm}
The energy can be obtained from the free energy as $E=\frac{\partial(\beta F)}{\partial\beta}$. There is a subtlety at the transition point, 
where $|u_1|$ can take any value between 0 and $\frac{1}{2}$. 
Hence we introduce a small interaction, which amounts to the following modification of the free energy: 
\begin{eqnarray}
\beta F 
=
\frac{DN^2\beta}{2}
+
N^2(1-De^{-\beta}+\epsilon_2)|u_1|^2
+
\epsilon_4 N^2|u_1|^4
+\cdots. 
\end{eqnarray}
Here $\epsilon_2$ and $\epsilon_4$ are functions of $\beta$, and the higher order terms represented by dots
will be negligible in the ensuing analysis. Let $T=T_c'$ be the solution of $f(\beta)\equiv 1-De^{-\beta}+\epsilon_2=0$, 
and introduce $\Delta\beta$ by $\Delta\beta=\beta-\frac{1}{T_c'}$. 

By solving the saddle point equation $\frac{\partial(\beta F)}{\partial |u_1|}=0$, 
the saddle point can be written as 
\begin{eqnarray}
|u_1|^2
\simeq
-\frac{f(\beta)}{2\epsilon_4}
\simeq
-\frac{f'(T_c'^{-1})\Delta\beta}{2\epsilon_4},  
\end{eqnarray} 
and the free energy becomes
\begin{eqnarray}
\beta F
=
\frac{DN^2\beta}{2}
-
\frac{N^2(f'(T_c'^{-1}))^2}{4\epsilon_4}(\Delta\beta)^2.
\end{eqnarray}  
Therefore, 
\begin{eqnarray}
E
&=&
\frac{DN^2}{2}
-
\frac{N^2(f'(T_c'^{-1}))^2}{2\epsilon_4}\Delta\beta
+
\frac{N^2(f'(T_c'^{-1}))^2\epsilon'_4}{4\epsilon_4^2}(\Delta\beta)^2
\nonumber\\
&\simeq&
\frac{DN^2}{2}
-
\frac{N^2(f'(T_c'^{-1}))^2}{2\epsilon_4}\Delta\beta
\nonumber\\
&\simeq&
\frac{DN^2}{2}+f'(T_c'^{-1})\cdot N^2P^2
\nonumber\\
&\to&
\frac{DN^2}{2}+N^2P^2. 
\qquad 
(\epsilon_2\to 0)
\end{eqnarray} 
Here we have assumed $\epsilon_4$ and $\epsilon_4'$ are of the same order, which is true in reasonable examples.
The final form does not depend on the detail of $\epsilon_2$ and $\epsilon_4$.

\subsection{The nonzero norm of $|E\rangle_{\rm inv}$}\label{sec:nonzero-norm}
\hspace{0.51cm}
In addition to $|E\rangle_{\rm inv}$, 
there is another canonical mapping between the $SU(M)$- and $SU(N)$-invariant states,
which can further be used to show that the state $\int dU {\cal U}\left(|E;{\rm SU}(M)\rangle\right) $ has non-zero norm.
Such canonical extension works as follows. 
We start with the state $|E,{\rm SU}(M)\rangle$, which as explained above has the general form
\begin{equation}
|E,{\rm SU}(M)\rangle
=
\sum
\sum_{\alpha',\beta',\ldots =1}^{M^2-1}
c_{IJ\cdots K\cdots L\cdots}{}^{\alpha'\beta'\cdots}
\hat{a}^\dagger_{I\alpha'}\hat{a}^\dagger_{J\beta'}\cdots|0\rangle\ .
\end{equation}
It is $\su{M}$-invariant since the adjoint indices $\alpha', \beta',\cdots$ are all contracted.
The coefficients $c_{IJ\cdots K\cdots L\cdots}{}^{\alpha'\beta'\cdots}$ is made of the Kronecker delta
and the structure constant of $\su{M}$, and transform covariantly under $\su{M}$. 
We can obtain the $\su{N}$ generalization of these coefficients by using the Kronecker delta and 
structure constant of $\su{N}$.  
Equivalently, we replace ${\rm Tr}_{M\times M}(\hat{a}_I\hat{a}_J\cdots)$ with ${\rm Tr}_{N\times N}(\hat{a}_I\hat{a}_J\cdots)$. 

Now the simple extension leads to an $\su{N}$-invariant state\footnote{
Note that this state is not an energy eigenstate in generic theories with interactions.  
}
 \begin{equation}
|E',{\rm SU}(N)\rangle
=
\sum
\sum_{\alpha,\beta,\ldots =1}^{N^2-1}
c_{IJ\cdots K\cdots L\cdots}{}^{\alpha\beta\cdots}
\hat{a}^\dagger_{I\alpha}\hat{a}^\dagger_{J\beta}\cdots|0\rangle\ .
\end{equation}
Further notice that since the $\alpha'$ and the $\alpha''$ sectors are orthogonal, the inner product $\langle E', {\rm SU}(N)|E,{\rm SU}(M)\rangle$
is automatically of the form of norm square since the $\alpha''$ sector operators in  $\langle E', {\rm SU}(N)|$, which enters in the sum,  move freely to the right and annihilate the ground state. 
This means
 \begin{equation}
\langle E', {\rm SU}(N)|E,{\rm SU}(M)\rangle
=
\langle E, {\rm SU}(M)|E,{\rm SU}(M)\rangle
>
0\ .  
\label{suN-SuM-inner-product}
\end{equation}

With this extension we can proceed to show that the state $\int dU {\cal U}\left(|E;{\rm SU}(M)\rangle\right)$  is non-zero. For this we consider
\begin{eqnarray}
\lefteqn{
\langle E',{\rm SU}(N)|\int dU\,{\cal U}\left(|E;{\rm SU}(M)\rangle\right)
}\nonumber\\
&=& 
\int dU \left(\langle E',{\rm SU}(N)|{\cal U}\right)\left(|E;{\rm SU}(M)\rangle\right)
\nonumber\\
&=& 
\left(\int dU\right)\cdot\langle E', {\rm SU}(N)|E;{\rm SU}(M)\rangle
\nonumber\\
&=& 
 {\text{ Vol}} ({\rm SU}(N))\cdot \langle E, {\rm SU}(M)|E,{\rm SU}(M)\rangle
>
 0. 
\end{eqnarray}
This means the left hand side is non-zero, which proves that $\int dU\,{\cal U}\left(|E;{\rm SU}(M)\rangle\right) $ cannot vanish.

Let us also confirm the linear independence. 
We take the orthonormal basis of the $\su{M}$ theory $|E, i,{\rm SU}(M)\rangle$, 
where $i$ is a label which distinguishes the states when the energy is degenerate. 
From this, we have two kinds of the $\su{N}$-extensions $|E, i\rangle_{\rm inv}$ and $|E, i,{\rm SU}(N)\rangle$ as explained above. 
These two extensions have a simple one-to-one correspondence because the inner product 
$\langle E, i,{\rm SU}(N)|E',j\rangle_{\rm inv}$ is nonzero only when $E=E'$ and $i=j$. 
The same relation also shows that the $|E, i\rangle_{\rm inv}$ are linearly independent.  

\section{Hamiltonian formulation on lattice}\label{sec:lattice_formulation}
\hspace{0.51cm}
Let us first recall the Hamiltonian formulation of lattice gauge theory by Kogut-Susskind \cite{Kogut:1974ag}
on 3d flat space.  Three spatial directions are discretized, while time is continuous. 
The Hamiltonian consists of the electric term $\hat{H}_{\rm E}$ and the magnetic term $\hat{H}_{\rm M}$ respectively, 
$\hat{H}=\hat{H}_{\rm E} + \hat{H}_{\rm M}$, 
where 
\begin{eqnarray}
\hat{H}_{\rm E}  = \frac{a^3}{2}\sum_{\vec{x}}\sum_{\mu}\sum_{\alpha=1}^{N^2-1}
\left(\hat{E}^\alpha_{\mu,\vec{x}}\right)^2
\end{eqnarray}
and  
\begin{eqnarray}
\hat{H}_{\rm M} = -\frac{1}{ag_{\rm YM}^2}\sum_{\vec{x}}
\sum_{\mu<\nu}
{\rm Tr} \left(\hat{U}_{\mu,\vec{x}}\hat{U}_{\nu,\vec{x}+\hat{\mu}}\hat{U}^\dagger_{\mu,\vec{x}+\hat{\nu}}\hat{U}^\dagger_{\nu,\vec{x}}\right).
\end{eqnarray}
Here $\hat{U}_\mu(\vec{x})$ is the unitary link variable connecting $\vec{x}$ and $\vec{x}+\hat{\mu}$, 
where $\hat{\mu}$ is the unit vector along the $\mu$-direction ($\mu=x,y$ or $z$), and $a$ is the lattice spacing.  
The commutation relations are given by\footnote{
 Intuitively, $\hat{U}_{\mu,\vec{x}}\simeq e^{iag_{\rm YM}\hat{A}_{\mu,\vec{x}}^\alpha\tau_\alpha}$ 
with the Hermitian gauge field $\hat{A}_{\mu,\vec{x}}$, and 
$[\hat{A}_{\mu,\vec{x}}^\alpha,\hat{E}_{\nu,\vec{y}}^\beta]=i\delta_{\mu\nu}\delta_{\vec{x}\vec{y}}\delta_{\alpha\beta}$.
 }
\begin{eqnarray}
& &
[\hat{E}^\alpha_{\mu,\vec{x}},\hat{U}_{\nu,\vec{y}}] = ag_{\rm YM}\delta_{\mu\nu}\delta_{\vec{x}\vec{y}}\cdot \tau^\alpha \hat{U}_{\nu,\vec{y}},  
\nonumber\\
& &
[\hat{E}_{\mu,\vec{x}},\hat{E}_{\nu,\vec{y}}] 
=
[\hat{U}_{\mu,\vec{x}},\hat{U}_{\nu,\vec{y}}] 
=
[\hat{U}_{\mu,\vec{x}},\hat{U}^\dagger_{\nu,\vec{y}}] 
=
0. 
\end{eqnarray}
Here
$\tau_\alpha$ ($\alpha=1,2,\cdots,N^2-1$) are generators of the SU$(N)$ algebra 
which satisfy ${\rm Tr}(\tau_\alpha\tau_\beta)=\delta_{\alpha\beta}$, 
$\sum_\alpha\tau_\alpha^{ij}\tau_\alpha^{kl}=\delta^{il}\delta^{jk}-\frac{\delta^{ij}\delta^{kl}}{N}$. 
The Hamiltonian described above corresponds to the $A_t=0$ gauge. 
Correspondingly, the gauge singlet constraint is imposed by hand, choosing
states to be gauge-invariant. 
This can be achieved by acting with Wilson loops 
$\hat{W}=Tr\left(\hat{U}_{\mu,\vec{x}}\hat{U}_{\nu,\vec{x}+\hat{\mu}}\cdots \hat{U}_{\rho,\vec{x}-\hat{\rho}}\right)$
on the vacuum. 
The inner product is defined by using the Haar measure on the group manifold. 

We separate $\hat{E}^\alpha_{\mu,\vec{x}}$ into the $\su{M}$ part $\hat{E}^{\alpha'}_{\mu,\vec{x}}$
and the rest $\hat{E}^{\alpha''}_{\mu,\vec{x}}$. 
Correspondingly, we can introduce $\hat{U}'_{\mu,\vec{x}}$ and $\hat{U}''_{\mu,\vec{x}}$ such that 
\begin{eqnarray}
& &
\hat{U}_{\mu,\vec{x}}=\hat{U}'_{\mu,\vec{x}}\hat{U}''_{\mu,\vec{x}}
+
O(a^2), 
\nonumber\\
& &
{}[\hat{E}^{\alpha'}_{\mu,\vec{x}},
\hat{U}'_{\nu,\vec{y}}]= ag_{\rm YM}\delta_{\mu\nu}\delta_{\vec{x}\vec{y}} \tau^{\alpha'} \hat{U}'_{\nu,\vec{y}}, 
\qquad
{}[\hat{E}^{\alpha'}_{\mu,\vec{x}},
\hat{U}''_{\nu,\vec{y}}]=0, 
\nonumber\\
&&
{}[\hat{E}^{\alpha''}_{\mu,\vec{x}},\hat{U}'_{\nu,\vec{y}}]=0, 
\qquad
{}[\hat{E}^{\alpha''}_{\mu,\vec{x}},
\hat{U}''_{\nu,\vec{y}}]= ag_{\rm YM}\delta_{\mu\nu}\delta_{\vec{x}\vec{y}} \tau^{\alpha''} \hat{U}''_{\nu,\vec{y}}, 
\nonumber\\
& &
[\tau^{\alpha''},
\hat{U}'_{\mu,\vec{x}}]
=
0, 
\qquad
[\tau^{\alpha'},
\hat{U}''_{\mu,\vec{x}}]
=
0. 
\end{eqnarray}

The states in the $\su{M}$ sector $|E;{\rm SU}(M)\rangle$
can be obtained by acting with the loop consisting of $\hat{U}'$
on the gauge-invariant vacuum, which in this case corresponds to the Fock vacuum of the Kaluza-Klein modes. 
(As $g_{\rm YM}$ becomes smaller, we can take such states parametrically close to the energy eigenstates of the full Hamiltonian.)
The SU($N$) transformation on $U'$ and $U''$ can be defined in a straightforward manner, and, by using them, also $|E\rangle_{\rm inv}$.  

Another SU($N$)-extension $|E';{\rm SU}(N)\rangle$ can be defined by replacing $U'$'s in $|E;{\rm SU}(M)\rangle$ with $U$'s. 
Through it, we can derive a relation analogous to \eqref{suN-SuM-inner-product}. 

The Hamiltonian given above applies to flat space, including the torus compactification. 
In principle, the compactification on the sphere can be achieved as follows.\footnote{
This is rarely done because the parameter fine tuning needed for achieving the desired continuum limit
is technically very difficult. 
} 
Firstly, we make a lattice with the topology of sphere. 
Locally, the plaquette $\hat{U}_{\mu,\vec{x}}\hat{U}_{\nu,\vec{x}+\hat{\mu}}\hat{U}^\dagger_{\mu,\vec{x}+\hat{\nu}}\hat{U}^\dagger_{\nu,\vec{x}}$ is identified with $e^{ia^2g_{\rm YM}\hat{F}_{\mu\nu}}$, where $\hat{F}_{\mu\nu}$ is the field strength. 
Therefore, the magnetic term on the sphere can be obtained by 
\begin{eqnarray}
\hat{H}_{\rm M} 
&=&
 -\frac{1}{4ag_{\rm YM}^2}\sum_{\vec{x}}
\sum_{\mu,\nu,\rho,\sigma}
\sqrt{-g(\vec{x})}
g^{\mu\rho}(\vec{x})
g^{\nu\sigma}(\vec{x})
\nonumber\\
& &
\times
{\rm Tr}\left[
\left(
1-
 (\hat{U}_{\mu,\vec{x}}\hat{U}_{\rho,\vec{x}+\hat{\mu}}\hat{U}^\dagger_{\mu,\vec{x}+\hat{\rho}}\hat{U}^\dagger_{\rho,\vec{x}})
\right)
\left(
1-
 (\hat{U}_{\nu,\vec{x}}\hat{U}_{\sigma,\vec{x}+\hat{\nu}}\hat{U}^\dagger_{\nu,\vec{x}+\hat{\sigma}}\hat{U}^\dagger_{\sigma,\vec{x}})
\right)
\right]. 
\end{eqnarray}
Here we have assumed the lattice with the topology of the sphere, and choose the metric $g^{\mu\rho}(\vec{x})$ appropriately
so that the sphere is actually realized.  
The electric term is 
\begin{eqnarray}
\hat{H}_{\rm E}  = \frac{a^3}{2}\sum_{\vec{x}}
\sqrt{-g(\vec{x})}
\sum_{\mu}\sum_{\alpha=1}^{N^2-1}
\left(\hat{E}^\alpha_{\mu,\vec{x}}\right)^2. 
\end{eqnarray}

\section{More on the properties of partial deconfinement }\label{sec:super_renormalizable}
\hspace{0.51cm}
In this appendix we consider the cases in which the assumptions made in Sec.\ref{sec:Partial_deconfinement_review} can fail. 
Probably the simplest example is the bosonic Yang-Mills matrix model (dimensional reduction of pure Yang-Mills to $(0+1)$ dimension). In this theory, the energy scale is determined by the 't Hooft coupling $\lambda=g_{\rm YM}^2N$, which has the dimension of $({\rm mass})^3$. 
Based on dimensional counting, the deconfinement temperature (from confinement to partial deconfinement) is proportional to 
$\lambda^{1/3}$, and if we naively truncate 
the theory to $\su{M}$, then it changes to $(g_{\rm YM}^2M)^{1/3}=\lambda^{1/3}(M/N)^{1/3}$. Clearly, a naive truncation is not working. 
Another basic example is three-dimensional pure Yang-Mills on the flat noncompact space. 
The energy scale is set by the 't Hooft coupling $\lambda=g_{\rm YM}^2N$, which has the dimension of mass, 
and the deconfinement temperature scales as $g_{\rm YM}^2M=\lambda\times\frac{M}{N}$. 
Similar complication can arise when the coupling runs with the energy scale, such as in QCD. 

In these cases, if we assume that the energy is described by the zero-point energy on top of the $M^2$ units of the excitations,  
\begin{eqnarray}
E= aN^2+b(T)M^2, 
\end{eqnarray}
would be a natural relation. Note that $M$ is a function of $T$. 
Based on the numerical simulation on the lattice, we know that the transition takes place in a narrow temperature range. 
Therefore the temperature dependence of $b(T)$ in the transition region can be neglected. 
In that case the situation is close to the free theories studied in Sec.~\ref{sec:matrix_model} and Sec.~\ref{sec:4d_YM}.
The condition for the Polyakov loop phases \eqref{eq:Polyakov_loop} is expected by the same assumption. 
For the bosonic matrix model, the result of the numerical simulation \cite{Bergner:2019rca}
reproduces these relations rather precisely.

\bibliographystyle{utphys}
\bibliography{partial-deconfinement-examples}
\end{document}